\documentclass[12pt]{article}
\usepackage{a4}
\usepackage{amssymb}
\usepackage{cite}
\newcommand{\be}{\begin{equation}}
\newcommand{\ee}{\end{equation}}
\newcommand{\bea}{\begin{eqnarray}}
\newcommand{\eea}{\end{eqnarray}}

\begin{document}
\thispagestyle{empty}
\def\thefootnote{\fnsymbol{footnote}}
%\begin{flushright}
%{\sf ZMP-HH/09-17}\\%[1mm]
%{\sf Hamburger$\;$Beitr\"age$\;$zur$\;$Mathematik$\;$Nr.$\;$ 345}
%     \vskip 2em
%\end{flushright}
%\vskip 2.0em
\vspace*{1cm}
\begin{center}\Large
Defects in $G/H$ coset, $G/G$ topological field theory and discrete Fourier-Mukai transform.
\end{center}\vskip 1.5em
\begin{center}
Gor Sarkissian
\footnote{\scriptsize
~Email address: \\
$~$\hspace*{2.4em} gor@imf.au.dk
}
\end{center}
\begin{center}
Department of Mathematical Sciences, \ University of \AA rhus,\\
Ny Munkegade 118, \ DK\,--\,8000\, \AA rhus C
\end{center}
\vskip 1.5em
\begin{center} June 2010 \end{center}
\vskip 2em
\begin{abstract} \noindent
In this paper we  construct defects in coset $G/H$  theory. Canonical quantization  of  the gauged WZW model $G/H$ with $N$ defects
on a cylinder and a strip is performed and the symplectomorphisms between the corresponding phase spaces
and those of double Chern-Simons theory on an annulus and a disc with Wilson lines are established.
Special attention to topological coset $G/G$ has been paid. We prove that a $G/G$ theory on a cylinder with $N$ defects
coincides with Chern-Simons theory on a torus times the time-line $R$ with $2N$ Wilson lines. We have shown also that a $G/G$ theory on a strip 
with $N$ defects coincides with Chern-Simons theory on a sphere times the time-line $R$ with $2N+4$ Wilson lines.
This particular example of topological field theory enables us to penetrate into a general picture of defects 
in semisimple 2D topological field theory. We conjecture that defects in this case described by a 2-category of matrices of vector spaces
and that the action of defects on boundary states is given by the discrete Fourier-Mukai transform.
 \end{abstract}
\vspace{0.5cm}

\setcounter{footnote}{0}
\def\thefootnote{\arabic{footnote}}
\newpage
%%%%%%%%%%%%%%%%%%%%%%%%%%%%%%%%%%%%%%%%%%%%%%%%%%%%%%%%%%%%%%%%%%%%%%%%

\section{Introduction}
In this paper we study defects in the gauged WZW model.  We construct the phase spaces of the gauged WZW model
in the presence of defects and relate them to the moduli spaces of flat connections on punctured Riemann surfaces.
We generalize here the results of the paper  \cite{Sarkissian:2009hy} on defects in the WZW model to the coset model, 
using canonical quantization technique of the gauged WZW model developed in 
 \cite{Gawedzki:2001ye}.
Let us briefly remind the principal results of these papers.
 
In the paper \cite{Sarkissian:2009hy}, canonical quantization of the WZW model with defects has been performed.
Using the Lagrangian formulation of the WZW model with defects and boundaries
the following symplectomorphisms have been established:
\begin{enumerate}
\item
The symplectic phase space of the WZW model with $N$ defects on a cylinder is isomorphic
to that of a Chern-Simons theory on the annulus ${\cal A}$ times the time-line $R$ with $N$ time-like Wilson lines.
\item
The symplectic phase space of the WZW model with $N$ defects on a strip is isomorphic
to that of Chern-Simons theory on a disc $D$ times the time-line $R$ with $N+2$ time-like Wilson lines.

In the paper \cite{Gawedzki:2001ye},  using the Lagrangian formulation of the gauged WZW $G/H$ model on a cylinder and on a strip
the following symplectomorphisms have been established:

\item
The phase space of the gauged WZW $G/H$ model on a cylinder is isomorphic to the phase space 
of the double Chern-Simons theory on the annulus ${\cal A}$ times the  time-line $R$.

\item
The phase space of the gauged WZW $G/H$ model on a strip is isomorphic to the phase space 
of the double Chern-Simons theory on $D\times R$  with $G$ and $H$ gauge fields both
coupled to two time-like Wilson lines.

In the special case of topological coset $G/G$  these isomorphisms take the form:

\item
The phase space of the gauged WZW $G/G$ model on a cylinder is isomorphic to the phase space 
of the Chern-Simons theory on the torus $T^2={\cal A}\cup (-{\cal A})$ times the time-line $R$.

\item
The phase space of the gauged WZW $G/G$ model on a strip is isomorphic to the phase space 
of the Chern-Simons theory on the sphere $S^2=D\cup (-D)$  times $R$ with
four time-like Wilson lines.

We show here that the combined versions of the symplectomorphisms  1-2 and 3-4 for the gauged WZW model in the presence 
of defects take the form:

\item
The phase space of the gauged WZW $G/H$ model on a cylinder  with $N$ defects is isomorphic to the phase space 
of the double Chern-Simons theory on ${\cal A}\times R$ with $G$ and $H$ gauge fields both coupled to 
$N$ Wilson lines.

\item
The phase space of the gauged WZW $G/H$ model on a strip  with $N$ defects is isomorphic to the phase space 
of the double Chern-Simons theory on $D\times R$  with $G$ and $H$ gauge fields both
coupled to $N+2$ time-like Wilson lines.

In the special case of the topological coset $G/G$  these isomorphisms take the form:

\item
The phase space of the gauged WZW $G/G$ model on a cylinder with $N$ defects is isomorphic to the phase space 
of the Chern-Simons theory on  $T^2\times R$ with $2N$ Wilson lines.

\item
The phase space of the gauged WZW $G/G$ model on a strip with $N$ defects  is isomorphic to the phase space 
of the Chern-Simons theory on $S^2\times R$ with
$2N+4$ time-like Wilson lines.
\end{enumerate}

The last two isomorphisms allow us to achieve a very detailed picture of defects
in this particular example of topological field theory.  This picture enables us to conjecture
that in general defects in semisimple 2D TFT should be described by means of 
a 2-category of matrices of vector spaces. Previously the relation of defects and 2-categories in conformal
field theory was discussed in \cite{Davydov:2010rm,Frohlich:2006ch,Schweigert:2006af}.

The paper is organised in the following way.

In the second section we review the canonical quantization of the WZW model. In the third section 
we review the canonical quantization of the gauged WZW model. In  section 4 we present
defects in gauged WZW model, perform canonical quantization and establish the isomorphism 7 in the list above.
 In  section 5 we consider defects in the gauged WZW model on a strip and establish the  isomorphism number 8.
In  section 6 we consider defects in the topological coset $G/G$, establish isomorphisms  9 and 10, and
make conjectures on defects properties in general 2D semisimple TFT.

\section{Bulk WZW model}
In this section we review the canonical quantization of the
WZW model with compact, simple, connected and simply connected group $G$  on the cylinder $\Sigma=R\times S^1=(t,x\; {\rm mod}\; 2\pi)$ 
\cite{Chu:1991pn,Falceto:1992bf,Gawedzki:1990jc}. 
The world-sheet action of the bulk WZW model is \cite{Witten:1983ar}
\bea
S^{\rm WZW}(g)&=&{k \over 4\pi}\int_{\Sigma}{\rm Tr}(g^{-1}\partial_{+} g)(g^{-1}\partial_{-}g)dx^+dx^-
 +{k \over 4\pi}\int_B {1\over 3}{\rm tr}(g^{-1}d g)^3\\ \nonumber
&\equiv& { k\over{4 \pi}} \left[ \int_{\Sigma}dx^+dx^- L^{\rm kin} 
+ \int_B \omega^{\rm WZ}\right] \, ,
 \eea
where $x^{\pm}=x\pm t$.
%\be
%\omega^{WZ}(g)={1\over 3}{\rm tr}(g^{-1}d g)^3
%\ee
The phase space of solutions ${\cal P}$ can be described by the Cauchy data 
\footnote{ Surely we can choose any time slice, but for simplicity we always below take the slice $t=0$.}
at $t=0$:
\be
g(x)=g(0,x)\;\;\; {\rm and}\;\;\; \xi_0(x)=g^{-1}\partial_tg(0,x)\, .
\ee

The corresponding symplectic form is \cite{Gawedzki:1990jc}:
\be
\Omega^{\rm bulk}={k\over 4\pi}\int_0^{2\pi}\Pi^{G}(g) dx\, ,
\ee
where 
\be\label{pigden}
\Pi^{G}(g)={\rm tr}\left(-\delta\xi_0g^{-1}\delta g+
(\xi_0+g^{-1}\partial_x g)(g^{-1}\delta g)^2\right)\, .
\ee
The $\delta$ denotes here the exterior derivative on the phase space ${\cal P}$.
It is easy to check that the symplectic form density $\Pi(g)$ has the following exterior derivative
\be\label{dpi}
\delta \Pi^{G}(g)=\partial_x\omega^{WZ}(g)\, ,
\ee
what implies closedness of the $\Omega$:
\be
\delta\Omega^{\rm bulk}=0\, .
\ee

The classical equations of motion are
\be\label{eqmot}
\partial_{-}J_L=0\;\;\;{\rm and} \;\;\;\partial_{+}J_R=0\, ,
\ee
where 
\be
J_L=-ik \partial_{+}gg^{-1}\;\;\;{\rm and}\;\;\; J_R=ikg^{-1}\partial_{-}g\, .
\ee
The general solution of (\ref{eqmot}) satisfying the boundary conditions:
\be
g(t,x+2\pi)=g(t,x)
\ee
is
\be\label{bulkdec}
g(t,x)=g_L(x^+)g^{-1}_R(x^-)
\ee
with $g_{L,R}$ satisfying the monodromy conditions:
\be\label{monl}
g_{L}(x^++2\pi)=g_{L}(x^+)\gamma\, ,
\ee
\be\label{monr}
g_{R}(x^-+2\pi)=g_{R}(x^-)\gamma
\ee
with the same matrix $\gamma$.
Expressing the symplectic form density $\Pi^{G}(g)$ in the terms of $g_{L,R}$ we obtain:
\be\label{denlr}
\Pi^G=
{\rm tr}\left[g_L^{-1}\delta g_L\partial_x(g_L^{-1}\delta g_L)-
g_R^{-1}\delta g_R\partial_x(g_R^{-1}\delta g_R)
+\partial_x (g_L^{-1}\delta g_Lg_R^{-1}\delta g_R)\right]\, .
\ee
Using (\ref{denlr}) and (\ref{monl}), (\ref{monr}) one derives for $\Omega^{\rm bulk}$:
\be\label{rldeco}
\Omega^{\rm bulk}=\Omega^{\rm chiral}(g_L,\gamma)-\Omega^{\rm chiral}(g_R, \gamma)\, ,
\ee
where
\be
\Omega^{\rm chiral}(g_L,\gamma)={k\over 4\pi}\int_0^{2\pi}{\rm tr}\left(g_L^{-1}\delta g_L\partial_x(g_L^{-1}\delta g_L)\right)dx+
{k\over 4\pi}{\rm tr}(g_L^{-1}\delta g_L(0)\delta\gamma\gamma^{-1})\, .
\ee
%and $\Omega_R$ is given by the same formula with $g_R\rightarrow g_L$.
The chiral field $g_L$ can be decomposed into the product of a closed loop in $G$,
a multivalued field in the Cartan subgroup and a constant element in $G$:
\be\label{decomp}
g_L(x^+)=h(x^+)e^{i\tau x^+/k}g_0^{-1}\, ,
\ee
where $h\in LG$, $\tau\in t$ ( the Cartan algebra) and $g_0\in G$. For the monodromy
of $g_L$ we find:
\be\label{mondecomp}
\gamma=g_0e^{2i\pi\tau/k }g_0^{-1}\, .
\ee
The parametrization (\ref{decomp}) induces the following decomposition of $\Omega^{\rm chiral}(g_L,\gamma)$:
\be\label{omdecom}
\Omega^{\rm chiral}(g_L,\gamma)=\Omega^{LG}(h,\tau)+{k\over 4\pi}\omega_{\tau}(\gamma)+{\rm tr}[(i\delta\tau)g_0^{-1}\delta g_0]\, ,
\ee
where $\Omega^{LG}(h,\tau)$ is :
\be\label{omlg}
\Omega^{LG}(h,\tau)={k\over 4\pi}\int_0^{2\pi}{\rm tr}[h^{-1}\delta h\partial_x(h^{-1}\delta h)+
{2i\over k}\tau(h^{-1}\delta h)^2
-{2i\over k}(\delta \tau)h^{-1}\delta h]dx
\ee
and $\omega_{\tau}(\gamma)$ is :
\be\label{omfik}
\omega_{\tau}(\gamma)={\rm tr}[g_0^{-1}\delta g_0e^{2i\pi\tau/k}g_0^{-1}\delta g_0e^{-2i\pi\tau/k}]\, .
\ee

Comparing (\ref{rldeco})  and (\ref{omlg}) to the formulae in appendix C  we see that the symplectic phase space of the 
WZW model on a cylinder coincides with that of Chern-Simons theory on the annulus times the time-line ${\cal A}\times R$.

\section{Gauged WZW model}
Here we review quantization of the gauged WZW model on the cylinder $\Sigma=R\times S^1=(t,x\; {\rm mod}\; 2\pi)$ 
as it is done in \cite{Gawedzki:2001ye}.

The action of the gauged WZW model is \cite{Bardakci:1987ee,Gawedzki:1988hq,Gawedzki:1988nj,Karabali:1988au}:
\be\label{gauact}
S^{G/H}=S^{\rm WZW}+S^{\rm gauge}\, ,
\ee
where
\be
S^{\rm gauge}={k\over 2\pi }\int_{\Sigma}L^{\rm gauge}\, ,
\ee
\be
L^{\rm gauge}(g,A)=-{\rm tr}[-\partial_{+}gg^{-1}A_{-}+g^{-1}\partial_{-}gA_{+}+gA_{+}g^{-1}A_{-}-A_{+}A_{-}]\, .
\ee
With the help of  the Polyakov-Wiegmann identities:
\begin{eqnarray}
\label{pwk}
L^{\rm kin}(g h) &=&  L^{\rm kin}(g) + L^{\rm kin} (h)
 +  {\rm Tr} \big(g^{-1}\partial_z g \partial_{\bar z} h h^{-1}\big)+
 {\rm Tr} \big(g^{-1} \partial_{\bar z} g\partial_z  h h^{-1}\big)\, , \\[1ex]
\label{pwwz}
\omega^{\rm WZ}(g h) &=& \omega^{\rm WZ}(g) + \omega^{\rm WZ}(h)
 - {\rm d}\Big({\rm Tr} \big(g^{-1} {\rm d}g  {\rm d}h h^{-1}\big)\Big)\, ,
\end{eqnarray}
it is easy to check that the action (\ref{gauact}) is invariant under the gauge transformation:
\be
g\rightarrow hgh^{-1}\, ,\hspace{1cm} A\rightarrow hAh^{-1}-dhh^{-1}
\ee
for $h: \Sigma\rightarrow H$.

The equations of  motions are:

\be
\label{eom}
D_{+}(g^{-1}D_{-}g)=0\, ,\;\;\:  {\rm Tr}(g^{-1}D_{-}gT_H)={\rm Tr}(gD_{+}g^{-1}T_H)=0\, , \;\;\; F(A)=0\, ,
\ee
where $D_{\pm}g=\partial_{\pm}g+[A_{\pm},\, g]$  and $T_H$  is any element in the $H$ Lie algebra.

The flat gauge field  $A$ can be written as $h^{-1}dh$ for $h:\, R^2\rightarrow H$ and satisfying:
\be
h(t, x+2\pi)=\rho^{-1}h(t,x)
\ee
for some $\rho\in H$.

Define $\tilde{g}=hgh^{-1}$. Note that $\tilde{g}$ satisfies
\be\label{mongrho}
\tilde{g}(t,x+2\pi)=\rho^{-1}\tilde{g}(t,x)\rho\, .
\ee

In the terms of $\tilde{g}$
 equations (\ref{eom})  take the form:
\be\label{eom2}
\partial_{+}(\tilde{g}^{-1}\partial_{-}\tilde{g})=0\, ,\;\;\; {\rm Tr}(\tilde{g}^{-1}\partial_{-}\tilde{g}T_H)={\rm Tr}(\tilde{g}\partial_{+}\tilde{g}^{-1}T_H)=0\, .
\ee
%where $\tilde{g}=hgh^{-1}$.

The canonical symplectic form density, obtained  following the general prescription \cite{Crnkovic:1987tz,Crnkovic:1986ex,Gawedzki:1990jc},
is given by:
\be\label{densgh}
\Pi^{G/H}(g,h)=\Pi^G(\tilde{g})+\partial_x\Psi(h,g)\, ,
\ee

where
\be\label{formim}
\Psi(h,g)={\rm tr}\; h^{-1}d h(g^{-1}d g+d g g^{-1}+g^{-1}h^{-1}d h g)\, .
\ee

Some properties of the form (\ref{formim}) are summarized in appendix A.

Integrating   (\ref{densgh})  we get  the canonical symplectic form:
%\be
%\Omega^{G/H}=\int_0^{2\pi}\Pi^G(\tilde{g})+\Psi(h(2\pi),g(2\pi))-\Psi(h(0),g(0))
%\ee

%Using equation (\ref{om1}) in appendix  A we get
\be\label{ghfor}
\Omega^{G/H}={k\over 4\pi}\int_0^{2\pi}\Pi^G(\tilde{g})dx+{k\over 4\pi}\Psi(\rho^{-1}, hgh^{-1}(0))\, .
\ee
Collecting (\ref{dpi}), (\ref{mongrho}) and (\ref{om4}) one can show that the form (\ref{ghfor}) is closed.

Equations (\ref{eom2})  can be solved  in the terms of the chiral fields:
\be\label{newdec}
\tilde{g}=g_{L}(x^+)g_{R}^{-1}(x^-)\, ,\hspace{1cm} {\rm Tr}(\partial_yg_Lg_L^{-1}T_H)={\rm Tr}(\partial_yg_Rg_R^{-1}T_H)=0
\ee
with the monodromy properties:
\be\label{monpr}
g_L(y+2\pi)=\rho^{-1}g_L(y)\gamma\, ,  \hspace{1cm} g_R(y+2\pi)=\rho^{-1}g_R(y)\gamma\, .
\ee

The monodromy properties (\ref{monpr}) imply  that the chiral fields $g_{L,R}$ should be written as products of fields as well:

\be\label{decomppp}
g_{L}=h_B^{-1}g_A\, , \hspace{1cm}  g_{R}=h_D^{-1}g_C\, ,
\ee
where $h_B, h_D\in H$ and $g_A,g_C\in G$.
The fields in (\ref{decomppp}) should additionally satisfy\footnote{One can arrive at the decomposition (\ref{decomppp}) with the properties (\ref{monhgl}) and (\ref{monhgr}) in the following way: 
taking, say,  a field $h_B$ satisfying the first part of (\ref{monhgl}), one can then define $g_A$ as $g_A\equiv h_Bg_L$, satisfying the second part of (\ref{monhgl}).}:
\be\label{idenpr}
{\rm tr} [T_H(\partial_y h_B h_B^{-1}-\partial_y g_A g_A^{-1})]=0\, ,\hspace{1cm}{\rm tr} [T_H(\partial_y h_D h_D^{-1}-\partial_y g_C g_C^{-1})]=0
\ee
and
\be\label{monhgl}
h_B(y+2\pi)=h_B(y)\rho\, ,\hspace{1cm}g_A(y+2\pi)=g_A(y)\gamma\, ,
\ee
\be\label{monhgr}
h_D(y+2\pi)=h_D(y)\rho\, ,\hspace{1cm}g_C(y+2\pi)=g_C(y)\gamma\, .
\ee

Using (\ref{idenpr}) one can show:
\be\label{id33}
{\rm tr}[g_{L}^{-1}\delta g_{L}\partial_y (g_{L}^{-1}\delta g_{L})]={\rm tr}[ g_{A}^{-1}\delta g_{A}\partial_y (g_{A}^{-1}\delta g_{A})
- h_{B}^{-1}\delta h_{B}\partial_y (h_{B}^{-1}\delta h_{B})+\partial_y(\delta h_B h_B^{-1}\delta g_A g_A^{-1})]
\ee
and similarly for $g_R$ and $h_D$, $g_C$.

Collecting (\ref{newdec}), (\ref{monpr}), (\ref{decomppp}), (\ref{monhgl}), (\ref{monhgr}), (\ref{id33}) and  (\ref{denlr})
one can show that
\be\label{gaugsymp}
\Omega^{G/H}=\Omega^{\rm chiral}(g_A,\gamma)-\Omega^{\rm chiral}(g_C,\gamma)-\Omega^{\rm chiral}(h_B,\rho)+\Omega^{\rm chiral}(h_D,\rho)
\ee

Comparing (\ref{gaugsymp}) with (\ref{rldeco}) and remembering that the latter is the symplectic form 
of the Chern-Simons theory on ${\cal A}\times R$, we arrive at the conclusion that the phase space of
the gauged WZW model on a cylinder coincides with that of double Chern-Simons theory \cite{Gawedzki:2001ye,Moore:1989yh} on  ${\cal A}\times R$.

\section{Defects in the gauged WZW model G/H}

Let us assume that one has a defect line $S$ separating the world-sheet into two
regions $\Sigma_1$ and $\Sigma_2$. In such a situation the WZW model is
defined by pair of maps $g_1$ and $g_2$.
On the defect line itself one has to impose conditions that relate the two maps. 
The necessary data are captured by the geometrical structure of a bibrane:
a bibrane is in particular a submanifold of the Cartesian product of the group G with itself
: $Q\subset G\times G$. The pair of maps $(g_1,g_2)$  are restricted by the requirement 
that the combined  map 
\be\label{defline}
S\rightarrow (G\times G): s\rightarrow (g_1(s),g_2(s)\in Q
\ee
takes its value in the submanifold $Q$. 
Additionally one should require, that on the submanifold $Q$ 
a two-form  $\varpi(g_1,g_2)$ exists satisfying the relation 
\be\label{extdef}
d\varpi(g_1,g_2)=\omega^{WZ}(g_1)|_{Q}-\omega^{WZ}(g_2)|_{Q}\, .
\ee
To write the action of the WZW model with defect one  should 
introduce  an auxiliary disc $D$ satisfying the conditions:
\be
\partial B_1=\Sigma_1+D\;\; {\rm and}\;\; \partial B_2=\Sigma_2+\bar{D}\, ,
\ee
and continue the fields $g_1$ and $g_2$ on this disc always holding 
the condition (\ref{defline}).
After this preparations the topological part of the action takes the form \cite{Fuchs:2007fw}:
\be\label{actdef}
S^{\rm top-def}={k \over 4\pi}\int_{B_1} \omega^{WZ}(g_1)+{k \over 4\pi}\int_{B_2} \omega^{WZ}(g_2)-{k\over 4\pi}\int_D\varpi(g_1,g_2)\, .
\ee

Equation (\ref{extdef}) guarantees that  (\ref{actdef}) is well defined.

Denote by $C_G^{\mu}$ a conjugacy class in group G:
\be
C_G^{\mu}=\{\beta f_{\mu}\beta^{-1}=\beta e^{2i\pi\mu/k}\beta^{-1},\;\;\;\; \beta\in G\}\, ,
\ee
where $\mu\equiv${\boldmath $\mu\cdot H$} is a highest weight representation integrable at level $k$,
taking value in the Cartan subalgebra of the G Lie algebra. 

To construct the gauged WZW model with defects we take for bibrane the following ansatz:
\be\label{bibrane}
(g_1,g_2)=(C_2C_1p, LpL^{-1})\, .
\ee
Here $C_1\in C_G^{\mu_1}$, $C_2\in C_H^{\mu_2}$, $p\in G$  and $L\in H$.

The Polyakov-Wiegmann identity (\ref{pwwz}) implies that the bibrane (\ref{bibrane}) satisfies the condition (\ref{extdef}) with the following  $\varpi$:
\be
\varpi(L,p,C_2,C_1)=\Omega^{(2)}(C_2,C_1)-{\rm tr}((C_2C_1)^{-1}d(C_2C_1)dpp^{-1})+\Psi(L,p)\, ,
\ee
where
\be\label{ombig}
\Omega^{(2)}(C_2,C_1)=\omega_{\mu_2}(C_2)-{\rm tr}(C_2^{-1}dC_2dC_1C_1^{-1})+\omega_{\mu_1}(C_1)
\ee
and $\omega_{\mu}(C)$ is defined in (\ref{omfik}) and $\Psi(L,p)$ is defined in (\ref{formim}).

In the following we show that this bibrane  provides a geometric realization of the Cardy defect \cite{Petkova:2000ip} corresponding to the primary $(\mu_1,\mu_2)$.

Some comments are in order at this point:
\begin{enumerate}
\item
In fact this is a folded version of the permutation brane for the gauged WZW model 
suggested in \cite{Sarkissian:2006xp}, and can be derived following the same lines as  for the permutation branes.

\item
Recall that primaries of the coset $G/H$ in the presence of the common center $C$ of the $G$ and $H$ are given by the pairs of the $G$ and $H$ primaries up to the 
field identification and selection rules \cite{DiFrancesco:1997nk,Gepner:1989jq,Gepner:1988wi}.
It was shown in  \cite{Elitzur:2001qd,Gawedzki:2001ye} that the product of the conjugacy classes $C_2C_1$ provides geometric realization of the
field identification and selection rules. Remind briefly the arguments. Field identification  follows immediately from the fact that
given an element of the common center $z$ one has the same object for $( f_{\mu_1}, f_{\mu_2})$ and    $(z f_{\mu_1},z^{-1} f_{\mu_2})$.
Selection rule is a consequence of the global issues \cite{Alekseev:1998mc,Fuchs:2007fw,Gawedzki:1999bq,Klimcik:1996hp}. Recall that when the second homotopy group of the bibrane worldvolume
is not trivial  the action is defined  up to the multiples of $2\pi$  only for the special values of $(\mu_1, \mu_2)$. In the absence
of the common center it is  enough to require that $(\mu_1,\mu_2)$ is a pair of $G$ and $H$ primaries.
The common center $C$ makes the gauge group $H/C$, and the above mentioned process of the continuation of the gauged transformed values of the $g_1$ and $g_2$ 
into the disc $D$  becomes non-trivial. Resolution of this problem leads to the selection rule,  requiring $\mu_2$ to lie somewhere in the integrable representation $\mu_1$, after some projection.\footnote{See 
\cite{DiFrancesco:1997nk} for details.}

%$k$({\boldmath $\mu_2$}$-${\boldmath $\mu_1$})$\cdot${\boldmath $\omega$}$\in 2\pi Z$,
%where {\boldmath $\omega$} is a common weight of $G$ and $H$.

\end{enumerate}

Now we can define the action of the gauged WZW model with defect as 
\be\label{actdefg}
S^{G/H-def}(g_1,g_2,A_1.A_2)=S^{\rm kin-def}(g_1,g_2)+S^{\rm gauge-def}(g_1,g_2,A_1,A_2)+S^{\rm top-def}\, ,
\ee

where
\be
S^{\rm kin-def}(g_1,g_2)={k \over 4\pi}\int_{\Sigma_1}L^{\rm kin}(g_1)dx^+dx^-+
{k \over 4\pi}\int_{\Sigma_2}L^{\rm kin}(g_2)dx^+dx^-
\ee
and
\be
S^{\rm gauge}(g_1,g_2,A_1,A_2)={k\over 2\pi }\int_{\Sigma_1}L^{\rm gauge}(g_1,A_1)
+{k\over 2\pi }\int_{\Sigma_2}L^{\rm gauge}(g_2,A_2)\, .
\ee
The gauge fields $A_1$ and $A_2$ are not restricted to the defect line.

One can check that the action (\ref{actdefg}) is invariant under the gauge transformations:
\bea\label{trang}
g_1&\rightarrow &h_1g_1h_1^{-1}\, ,\hspace{1cm} A_1\rightarrow h_1A_1h_1^{-1}-dh_1h_1^{-1}\, ,\\ \nonumber
g_2&\rightarrow &h_2g_2h_2^{-1}\, ,\hspace{1cm} A_2\rightarrow h_2A_2h_2^{-1}-dh_2h_2^{-1}\, ,
\eea
where $h_1: \Sigma\rightarrow H$, $h_2: \Sigma\rightarrow H$.
For this purpose note that under (\ref{trang}) the boundary parameters transform in the following way:
\be
 p\rightarrow  h_1ph_1^{-1}\, ,\;\;\; C_1\rightarrow  h_1C_1h_1^{-1}\, ,\;\;\;
C_2\rightarrow  h_1C_2h_1^{-1}\, ,\;\;\; L\rightarrow h_2Lh_1^{-1}\, .
\ee

The gauge invariance follows from the Polyakov-Wiegmann identities and the transformation properties 
of $\varpi(L,p,C_2,C_1)$:
\be\label{varptran}
\varpi(h_2Lh_1^{-1},h_1ph_1^{-1},h_1C_2h_1^{-1},h_1C_1h_1^{-1})-\varpi(L,p,C_2,C_1)\, ,
=-\Psi(h_1,C_2C_1p)+\Psi(h_2,LpL^{-1})
\ee
which can be obtained using formulae 
of appendix A.

Now we consider the gauged WZW model on the cylinder  $\Sigma=R\times S^1=(t,x\; {\rm mod}\; 2\pi)$ 
and put defect line at $x=a$ in parallel to the time line.

The variational equation $\delta S^{G/H-def}(g_1,g_2,A_1.A_2)=0$ implies the bulk equations (\ref{eom}) for $g_1,A_1$ and $g_2,A_2$
separately supplemented by the defect equations at $x=a$:
\be\label{dee1}
g_1^{-1}D_{-}g_1-L^{-1}g_2^{-1}D_{-}g_2L=0\, ,
\ee
\be\label{dee2}
C_2^{-1}g_1D_{+}g_1^{-1}C_2-L^{-1}g_2D_{+}g_2^{-1}L=0\, ,
\ee
\be\label{dee3}
L^{-1}D_tL=0\, ,\;\;\; C_2^{-1}D_tC_2=0\, ,
\ee
where $D_t=D_{+}-D_{-}$,   $D_{\pm}L=\partial_{\pm}L+A_{2\pm}L-LA_{1\pm}$, $D_{\pm}g_1=\partial_{\pm}g_1+[A_{1\pm},\, g_1]$,
$D_{\pm}g_2=\partial_{\pm}g_2+[A_{2\pm},\, g_2]$,  $D_{\pm}C_2=\partial_{\pm}C_2+[A_{1\pm},\, C_2]$.

Derivation of the equations (\ref{dee1}), (\ref{dee2}), (\ref{dee3}) is outlined in the appendix B.

Flat gauge fields can be parameterised as before:
\be
A_1=h_1^{-1}dh_1\, ,\hspace{1cm} A_2=h_2^{-1}dh_2\, .
\ee
Defining as before:
\bea
\tilde{g}_1=h_1g_1h_1^{-1}\, ,\hspace{1cm} \tilde{g}_2=h_2g_2h_2^{-1}\, ,\\ \nonumber
\tilde{C}_1=h_1C_1h_1^{-1}\, ,\hspace{1cm}\tilde{C}_2=h_1C_2h_1^{-1}\, ,\\ \nonumber
\tilde{p}=h_1ph_1^{-1}\, ,\hspace{1cm} \tilde{L}=h_2Lh_1^{-1}\, ,
\eea
we have the bulk equations (\ref{eom2}) for $\tilde{g}_1$ and $\tilde{g}_2$ and the defect equations  (\ref{dee1}), (\ref{dee2}), (\ref{dee3}) take the form:
\be
\tilde{g}_1^{-1}\partial_{-}\tilde{g}_1-\tilde{L}^{-1}\tilde{g}_2^{-1}\partial_{-}\tilde{g}_2\tilde{L}=0\, ,
\ee
\be
\tilde{C}_2^{-1}\tilde{g}_1\partial_{+}\tilde{g}_1^{-1}\tilde{C}_2-\tilde{L}^{-1}\tilde{g}_2\partial_{+}\tilde{g}_2^{-1}\tilde{L}=0\, ,
\ee
\be\label{eoml}
\tilde{L}^{-1}\partial_t\tilde{L}=0\, ,\;\;\; \tilde{C}_2^{-1}\partial_t\tilde{C}_2=0\, .
\ee
Equation (\ref{eoml}) implies that $\tilde{L}$ and $\tilde{C}_2$ are constant along the defect line.

Using that,  the bulk-defect equations can be solved in the terms of the chiral fields:
\be\label{1LGR}
\tilde{g}_1=g_{1L}g_{1R}^{-1}\, ,\hspace{1cm} {\rm Tr}(\partial_yg_{1L}g_{1L}^{-1}T_H)={\rm Tr}(\partial_yg_{1R}g_{1R}^{-1}T_H)=0\, ,
\ee
\be\label{2LGR}
\tilde{g}_2=g_{2L}g_{2R}^{-1}\, ,\hspace{1cm} {\rm Tr}(\partial_yg_{2L}g_{2L}^{-1}T_H)={\rm Tr}(\partial_yg_{2R}g_{2R}^{-1}T_H)=0\, ,
\ee

and

\be\label{LR12}
g_{2L}=\tilde{L}\tilde{C}_2^{-1}g_{1L}n^{-1}\, , \hspace{1cm} g_{2R}=\tilde{L}g_{1R}m^{-1}\, ,
\ee
with $m$ and $n$ $\in G$.
Equations (\ref{LR12}) imply
\be
(\tilde{g}_1(t,a), \tilde{g}_2(t,a))=(\tilde{C}_2\tilde{C}_1\tilde{p}, \tilde{L}\tilde{p}\tilde{L}^{-1})\, ,
\ee
where
\be\label{tilk}
\tilde{p}=\tilde{C}_2^{-1}g_{1L}(a+t)n^{-1}mg_{1R}^{-1}(a-t)\, ,
\ee
\be\label{tilc1}
\tilde{C}_1=\tilde{C}_2 ^{-1}g_{1L}(a+t)m^{-1}ng_{1L}^{-1}(a+t)\tilde{C}_2\, .
\ee
To have that $\tilde{C}_1\in C_G^{\mu_1}$ we should require that $d\equiv m^{-1}n\in C_G^{\mu_1}$.

Given that we consider GWZW model on a cylinder we should additionally require:
\be\label{mong12}
g_1(t,0)=g_2(t,2\pi)\, ,
\ee
\be\label{monh12}
h_1(t,0)=\rho h_2(t,2\pi)\, .
\ee

From (\ref{mong12}) and (\ref{monh12}) one obtains:
\be\label{g1g2rho}
\tilde{g}_1(t,0)=\rho\tilde{g}_2(t,2\pi)\rho^{-1}\, ,
\ee
and
\be\label{monprdef}
g_{1L}(y+2\pi)=\tilde{C}_2 \tilde{L}^{-1}\rho^{-1}g_{1L}(y)\gamma_L\, ,\hspace{1cm}
g_{1R}(y+2\pi)= \tilde{L}^{-1}\rho^{-1}g_{1R}(y)\gamma_R\, ,
\ee
with $\gamma_L$ and $\gamma_R$  satisfying the relation:
\be
\gamma_R^{-1}\gamma_L=d\, .
\ee
Comparing (\ref{monprdef}) with (\ref{monpr}) we see that the presence of the defect leads to the relative shifts between the left and right monodromies,
equal  to the defect conjugacy classes.

The monodromies (\ref{monprdef}) as before can be realized in the terms of the decomposition of the fields $g_{1L}$ and $g_{1R}$ 
as products:
\be
g_{1L}=h_B^{-1}g_A\, , \hspace{1cm}  g_{1R}=h_D^{-1}g_C
\ee
of the new fields $h_B$, $g_A$, $h_D$, $g_C$ possessing the monodromy properties:
\be
h_B(2\pi)=h_B(0)\rho\tilde{L}\tilde{C}_2^{-1}\, ,\hspace{1cm}
g_A(2\pi)=g_A(0)\gamma_L\, ,
\ee
\be
h_D(2\pi)=h_D(0)\rho\tilde{L}\, ,\hspace{1cm}
g_C(2\pi)=g_C(0)\gamma_R\, ,
\ee
and satisfying (\ref{idenpr}).

The symplectic form of the gauged WZW model with a defect can be written using the symplectic form density (\ref{densgh}) and the form $\varpi$:
\be\label{ghfrodef}
\Omega^{G/H-def}={k\over 4\pi}\left[\int_0^{a}\Pi^{G/H}(g_1,h_1)dx+\int_a^{2\pi}\Pi^{G/H}(g_2,h_2)dx-\varpi(g_1(a),g_2(a))\right]\, .
\ee
Substituting in (\ref{ghfrodef}) the symplectic form density (\ref{densgh})  and  using  the transformation property (\ref{varptran})
 we obtain:

\be\label{ghdef12}
\Omega^{G/H-def}={k\over 4\pi}\left[\int_0^{a}\Pi(\tilde{g}_1)dx+\int_a^{2\pi}\Pi(\tilde{g}_2)dx-\varpi( \tilde{L}, \tilde{p},  \tilde{C}_2, \tilde{C}_1)-\Psi(\rho,\tilde{g}_2(2\pi))\right]\, ,
\ee
where $\tilde{p}$ and $\tilde{C}_1$ defined in (\ref{tilk}) and (\ref{tilc1}).

Performing similar steps as before we arrive at the following expression for the symplectic form of the gauged WZW model with defects:

\bea
\Omega^{G/H-def}=\Omega^{\rm chiral}(g_A, \gamma_L)-\Omega^{\rm chiral}(g_C, \gamma_R)-
\Omega^{\rm chiral}(h_B, \rho\tilde{L}\tilde{C}_2^{-1})+\Omega^{\rm chiral}(h_D, \rho\tilde{L})+\\ \nonumber
{k\over 4\pi}\left[-\omega_{\mu_2}(\tilde{C}_2)-\omega_{\mu_1}(d)-{\rm tr}(d^{-1}\delta d\gamma_L^{-1}\delta\gamma_L)
-{\rm tr}(\tilde{C}_2^{-1}\delta \tilde{C}_2(\rho\tilde{L})^{-1}\delta(\rho\tilde{L}))\right]\, .
\eea

Recalling the decomposition (\ref{omdecom}) of $\Omega^{\rm chiral}$ and comparing with the corresponding
formulae in appendix C we arrive at the conclusion that the phase space of the gauged WZW model
on a cylinder with a defect line coincides with that of double Chern-Simons theory on ${\cal A}\times R$
with gauge fields of groups $G$ and $H$ coupled to a Wilson line. 
This result can be straightforwardly generalized to the presence of the $N$ defect lines.

\section{Defects in open coset model $G/H$}

Let us at the beginning  remind some facts on boundary coset model $G/H$  \cite{Elitzur:2001qd,Gawedzki:2001ye}.

Boundary condition corresponding to a  Cardy state $(\mu,\nu)$ is given by the product of the conjugacy classes
\be\label{ncond}
g|_{\rm boundary}=bc\, ,
\ee
where $b\in C_G^{\mu}$ and  $c\in C_H^{\nu}$.
As explained in the previous section in the presence of the common center $C$
$\mu$ and $\nu$ should satisfy the selection rule.

To write the action one should introduce an auxiliary disc $D$ satisfying the condition $\partial B=\Sigma+D$,
and continue the field $g$ on this disc, always taking value in product of conjugacy classes.

The action with the boundary conditions (\ref{ncond}) has the form:

\be\label{actghbndr}
S^{G/H-bndry}=S^{G/H}-{k\over 4\pi}\int_D \Omega^{(2)}(b,c)\, ,
\ee
where $ \Omega^{(2)}(b,c)$ is defined in (\ref{ombig}).

Consider a WZW model with a defect on the strip $R\times [0,\pi]$. Assume again that we have a defect at the point $x=a$
in parallel to the time line. The strip is divided into two parts with the fields $g_1,A_1$ and $g_2,A_2$.
We impose a Cardy boundary condition (\ref{ncond}) at $x=0$ on $g_1$ requiring:
\be\label{bcon0}
g_1(t,0)=C_3C_4\, , \hspace{1cm}C_3\in C_G^{\mu_3}\, ,\hspace{1cm}C_4\in C_H^{\mu_4}\, ,
\ee 
a defect condition (\ref{bibrane}) at $x=a$:
\be\label{bibraneaa}
(g_1,g_2)=(C_2C_1p, LpL^{-1})\, ,
\ee
and again a Cardy  boundary condition (\ref{ncond}) at $x=\pi$:
\be\label{bconpi}
g_2(t,\pi)=C_5C_6\, , \hspace{1cm}C_5\in C_G^{\mu_5}\, ,\hspace{1cm}C_6\in C_H^{\mu_6}\, .
\ee 
Let us analyze first the  consequences of the boundary condition (\ref{bcon0}) at the point $x=0$.

The boundary equations of motion resulting from the action (\ref{actghbndr}) at $x=0$ are derived in  \cite{Gawedzki:2001ye}:
\be\label{bndreom}
g_1^{-1}D_{-}g_1+C_4^{-1}g_1D_{+}g_1^{-1}C_4=0\, ,\hspace{1cm}  C_4^{-1}D_tC_4=0\, .
\ee
Representing  again the flat gauge field $A_1=h_1^{-1}dh_1$, and again defining $\tilde{g}_1=h_1g_1h_1^{-1}$,  $\tilde{C}_3=h_1C_3h_1^{-1}$,    $\tilde{C}_4=h_1C_4h_1^{-1}$ 
one can write (\ref{bndreom}) as:
\be\label{bnchir}
\tilde{g}_1^{-1}\partial_-\tilde{g}_1+\tilde{C}_4^{-1}\tilde{g}_1\partial_+\tilde{g}_1^{-1}\tilde{C}_4=0\, ,
\ee
\be\label{constc}
\tilde{C}_4^{-1}\partial_t\tilde{C}_4=0\, .
\ee

The last equation implies that $\tilde{C}_4$ is constant on the boundary.
Therefore using the chiral decomposition (\ref{1LGR})  $\tilde{g}_1=g_{1L} g_{1R}^{-1}$ one can solve (\ref{bnchir}):

\be\label{rrll}
g_{1R}(y)=\tilde{C}_4^{-1}g_{1L}(-y)R_0^{-1}
\ee
with $R_0\in G$.
Now we get that :
\be
\tilde{g}_1(t,0)=g_{1L}(t)R_0g_{1L}^{-1}(t)\tilde{C}_4\, .
\ee
The boundary condition (\ref{bcon0}) implies:
\be\label{bcon0t}
\tilde{g}_1(0,t)=\tilde{C}_3\tilde{C}_4\, ,\hspace{1cm}\tilde{C}_3\in C_G^{\mu_3}\, ,\hspace{1cm}\tilde{C}_4\in C_H^{\mu_4}\, .
\ee
We find that
\be\label{tc3}
\tilde{C}_3=g_{1L}(t)R_0g_{1L}^{-1}(t)\, .
\ee
To be in agreement with the requirement that $\tilde{C}_3\in C_G^{\mu_3}$  one should demand:
\be
R_0\in C_G^{\mu_3}\, .
\ee
The defect condition as before implies:
\be\label{LR123}
g_{1L}=\tilde{C}_2\tilde{L}^{-1}g_{2L}n\, , \hspace{1cm} g_{1R}=\tilde{L}^{-1}g_{2R}m\, ,
\ee
where $g_{2L}$, $g_{2R}$ are fields of the chiral decomposition (\ref{2LGR}): $\tilde{g}_2=g_{2L} g_{2R}^{-1}$.

From the boundary condition (\ref{bconpi}) we conclude:
\be\label{bconpit}
\tilde{g}_2(t,\pi)=\tilde{C}_5\tilde{C}_6\, , \hspace{1cm}\tilde{C}_5\in C_G^{\mu_5}\, ,\hspace{1cm}\tilde{C}_6\in C_H^{\mu_6}\, ,
\ee 
where $\tilde{C}_5=h_2C_5h_2^{-1}$,  $\tilde{C}_6=h_2C_6h_2^{-1}$.

To satisfy (\ref{bconpit})  we assume the following monodromy behaviour of $g_{1L}$:
\be\label{monodik}
g_{1L}(y+2\pi)=\rho^{-1}g_{1L}(y)\gamma\, .
\ee
From relations (\ref{rrll}), (\ref{LR123}) and (\ref{monodik}) we derive:
\be
\tilde{g}_2(t,\pi)=\tilde{L}\tilde{C}_2^{-1}g_{1L}(\pi +t)n^{-1}mR_0\gamma(\tilde{L}\tilde{C}_2^{-1}g_{1L}(\pi +t))^{-1}\tilde{L}\tilde{C}_2^{-1}
\rho^{-1}\tilde{C}_4\tilde{L}^{-1}\, .
\ee
We see that
\be\label{tc5}
\tilde{C}_5=\tilde{L}\tilde{C}_2^{-1}g_{1L}(\pi +t)n^{-1}mR_0\gamma(\tilde{L}\tilde{C}_2^{-1}g_{1L}(\pi +t))^{-1}\, ,
\ee and
\be\label{tc6}
\tilde{C}_6=\tilde{L}\tilde{C}_2^{-1}\rho^{-1}\tilde{C}_4\tilde{L}^{-1}\, .
\ee
To satisfy (\ref{bconpit}) we should demand:
\be
d^{-1}R_0\gamma=R_{\pi}\in C_G^{\mu_5}\, ,
\ee
\be
\tilde{C}_2^{-1}\rho^{-1}\tilde{C}_4=S_{\pi}\in C_H^{\mu_6}\, .
\ee
The symplectic form is
\be\label{omopen}
\Omega^{G/H-def-bndry}={k\over 4\pi}\left[\int_0^{a}\Pi(\tilde{g}_1)+\int_a^{\pi}\Pi(\tilde{g}_2)-\varpi( \tilde{L},\tilde{p}, \tilde{C}_2, \tilde{C}_1)
+\Omega(\tilde{C}_3,\tilde{C}_4)-\Omega(\tilde{C}_5,\tilde{C}_6)\right]\, .
\ee
In formula (\ref{omopen}) $\tilde{p}$, $\tilde{C}_1$, $\tilde{C}_3$, $\tilde{C}_5$,  $\tilde{C}_6$  are given by the equations 
(\ref{tilk}), (\ref{tilc1}), (\ref{tc3}), (\ref{tc5}), (\ref{tc6}) correspondingly.
Representing again
\be
g_{1L}=h_B^{-1}g_A\, ,
\ee
with $h_B$ and $g_A$ possessing the monodromy properties:
\be
h_B(y+2\pi)=h_B(y)\rho\, ,
\ee
\be
g_A(y+2\pi)=g_A(y)\gamma\, ,
\ee

and repeating the same steps as before we obtain:

\bea
{4\pi\over k} \Omega^{G/H-def-bndry}={4\pi\over k}\Omega^{\rm chiral}(g_A, \gamma)-
{4\pi\over k}\Omega^{\rm chiral}(h_B, \rho)+\omega_{\mu_3}(R_0)+\\ \nonumber
\omega_{\mu_4}(\tilde{C}_4)-\omega_{\mu_5}(R_{\pi})-\omega_{\mu_6}(S_{\pi})-
\omega_{\mu_2}(\tilde{C}_2)-\omega_{\mu_1}(d)\\ \nonumber
-{\rm tr}(R_0^{-1}\delta R_0\delta\gamma \gamma^{-1})+{\rm tr}(\delta d d^{-1}\delta R_0R_0^{-1})
+{\rm tr}(\delta dd^{-1}R_0\delta\gamma \gamma^{-1}R_0^{-1})\\ \nonumber
-{\rm tr}(\delta \tilde{C}_4\tilde{C}_4^{-1}\delta\rho\rho^{-1})-{\rm tr}(\delta \tilde{C}_2\tilde{C}_2^{-1}\rho^{-1}\delta\rho)
+{\rm tr}(\delta \tilde{C}_2\tilde{C}_2^{-1}\rho^{-1}\delta \tilde{C}_4\tilde{C}_4^{-1}\rho)\, .
\eea

Recalling again the decomposition (\ref{omdecom}) of $\Omega^{\rm chiral}$ and comparing with the corresponding
formulae in appendix C we arrive at the conclusion that the phase space of the gauged WZW model
on a strip with a defect line coincides with that of the double Chern-Simons theory on $D\times R$
with gauge fields of groups $G$ and $H$ coupled to three Wilson lines. 
This result can be straightforwardly generalized to the presence of the $N$ defect lines.

\section{Topological G/G coset}

\subsection{Bulk G/G coset}
In this section we consider the bulk $G/H$ model studied in section 2 for the special case $G=H$.
It was shown in section 2 that the phase space of the bulk $G/H$ model is symplectomorphic to that of the double Chern-Simons theory
on $R\times {\cal A}$. In the special case, when $G=H$ it becomes a Chern-Simons theory on the torus times $R$ :
 $R\times ({\cal A}\cup (-{\cal A}))=R\times T^2$. This result can be obtained also by a direct calculation.

In the case when $G=H$ the equations of motion (\ref{eom2}) imply  that  $\tilde{g}$ is $(t,x)$ independent and therefore 
the symplectic form  $\Omega^{G/H}$ (\ref{ghfor}) reduces to 
\be\label{omgg}
\Omega^{G/G}={k\over 4\pi}\Psi(\rho,\tilde{g}^{-1})\, .
\ee
The fact that $\tilde{g}$ is constant on a cylinder and the relation (\ref{mongrho}) also imply
\be\label{phspgg}
\rho\tilde{g}\rho^{-1}\tilde{g}^{-1}=I\, .
\ee
Comparing   (\ref{omgg}) and (\ref{phspgg}) with the corresponding formulae
reviewed in appendix C we arrive at the conclusion that the phase space of a bulk $G/G$ theory on a cylinder is symplectomorphic to that of
a Chern-Simons theory on  $T^2\times R$. 
The quantization of the latter gives rise to the space of the 0-point conformal blocks of the WZW theory on the torus.
The dimension of the space of conformal blocks on a  Riemann surface of genus $g$ with insertion of the primary fields with labels
$\mu_n$ is:
\be\label{dimcoblock}
N_{\mu_n}(g)=\sum_{\alpha}(S_0^{\alpha})^{2-2g}\prod_n(S^{\alpha}_{\mu_n}/S^{\alpha}_0)\, .
\ee
This implies that the Hilbert space of the quantized $G/G$ theory on a cylinder
has dimension equals to the number of the integrable primaries.
The equivalence of the topological $G/G$ coset  on a cylinder $R\times S^1$ with a Chern-Simons on $R\times T^2$ demonstrated here 
is actually a particular case of the more general equivalence of the topological $G/G$ coset on a Riemann surface $\Sigma$
and the Chern-Simons theory on $\Sigma \times S^1$ established in \cite{Blau:1993tv,Spiegelglas:1992jg,Witten:1991mm}.

%\subsection{Open topological G/G coset}

\subsection{A defect in a closed topological model G/G}
We have established in  section 3 that the phase space of the coset $G/H$ on a cylinder with a defect is symplectomorphic 
to that of a double Chern-Simons  theory on  $R\times {\cal A}$ with $G$ and $H$ gauge fields both coupled to a time like Wilson line.
In the case when $G=H$ we again arrive at the conclusion that the  topological coset $G/G$ on a cylinder with a defect line
is equivalent to the Chern-Simons theory on $R\times T^2$ with two time like Wilson lines. This again can be verified by a direct calculation.
For the case $G=H$ the bulk equations of motion imply that $\tilde{g}_1$ and  $\tilde{g}_2$ are $(t,x)$ independent.

Therefore one has:
\be\label{tilg12}
\tilde{g}_1(0)=\tilde{g}_1(a)=\tilde{C}_2\tilde{C}_1\tilde{p}\, ,\hspace{1cm} \tilde{L}\tilde{p}\tilde{L}^{-1}=\tilde{g}_2(a)=\tilde{g}_2(2\pi)\, .
\ee
From (\ref{g1g2rho}) we also obtain:
\be\label{tilgrho}
\tilde{g}_1(0)=\rho\tilde{g}_2(2\pi)\rho^{-1}\, .
\ee
Inserting (\ref{tilgrho}) in (\ref{tilg12})  we  get:
\be\label{defrel}
\tilde{C}_2\tilde{C}_1\tilde{p}=\rho\tilde{L}\tilde{p}\tilde{L}^{-1}\rho^{-1}\, .
\ee
The symplectic form (\ref{ghdef12}) now takes the form:
\be\label{ggtopdef}
\Omega^{G/G-def}=-{k\over 4\pi}\varpi(\rho\tilde{L},\tilde{p}, \tilde{C}_2, \tilde{C}_1)\, .
\ee
Comparing (\ref{defrel})  and (\ref{ggtopdef}) with the corresponding formulae in appendix C we arrive
at the conclusion that the topological coset $G/G$ on a cylinder with a defect line is symplectomorphic with 
that of a Chern-Simons theory on $T^2\times R$ with two Wilson lines.
The quantization of the latter gives rise to the space of the 2-point conformal blocks of the WZW theory on a torus.
Using equation (\ref{dimcoblock}) we can compute the dimension of the Hilbert space of  the quantized 
topological coset $G/G$ on cylinder with a defect line $(\mu_1,\mu_2)$ :
\be\label{dimdefcyl}
{\rm dim}\, H_{d_{\mu_1,\mu_2}}=\sum_{\alpha\beta}N_{\alpha\mu_1}^{\beta}N_{\beta\mu_2}^{\alpha}\, .
\ee
\subsection{Defects in the open topological model G/G}

Previously we have seen that the phase space of $G/H$ coset on a strip with a defect
is symplectomorphic to that of the double Chern-Simons theory on $D\times R$ with gauge fields $G$ and $H$ both
coupled to three Wilson lines. In the case when $G=H$ we arrive at the conclusion that the $G/G$ topological coset
on a strip with a defect line is equivalent to the Chern-Simons theory on sphere times $R$ : $(D\cup(-D))\times R=S^2\times R$  with six time-like Wilson lines.
This can be verified also directly.
In this case $\tilde{g}_1$ and  $\tilde{g}_2$ are $(t,x)$ independent and therefore one has:
\be\label{g34}
\tilde{g}_1(0)=\tilde{C}_3\tilde{C}_4=\tilde{C}_2\tilde{C}_1\tilde{p}=\tilde{g}_1(a)\, ,
\ee
\be\label{g12}
\tilde{g}_2(a)=\tilde{L}\tilde{p}\tilde{L}^{-1}=\tilde{C}_5\tilde{C}_6=\tilde{g}_2(2\pi)\, .
\ee
From equations (\ref{g34}) and (\ref{g12}) one obtains:
\be\label{sphersix}
(\tilde{L}\tilde{C}_1^{-1}\tilde{L}^{-1})(\tilde{L}\tilde{C}_2^{-1}
\tilde{L}^{-1})(\tilde{L}\tilde{C}_3\tilde{L}^{-1})(\tilde{L}\tilde{C}_4\tilde{L}^{-1})\tilde{C}_6^{-1}\tilde{C}_5^{-1}=I\, ,
\ee
and from (\ref{omopen}) one derives:
\be\label{omsix}
\Omega^{G/G-def-bndry}=-{k\over 4\pi}\varpi(\tilde{L}, \tilde{p}, \tilde{C}_2, \tilde{C}_1)
+{k\over 4\pi}\Omega(\tilde{C}_3,\tilde{C}_4)-{k\over 4\pi}\Omega(\tilde{C}_5,\tilde{C}_6)\, .
\ee
Comparing  (\ref{sphersix}) and (\ref{omsix})  with the corresponding equations in appendix C we arrive at the mentioned
symplectomorphism of the phase space of $G/G$ topological coset on a strip with a defect and a Chern-Simons theory on $S^2\times R$ with six Wilson lines.
The quantization of the latter gives rise to the space of the 6-point conformal blocks of the WZW theory on a sphere.
Using equation (\ref{dimcoblock}) we can compute the dimension of the Hilbert space of  the quantized 
topological coset $G/G$ on a strip with a defect line:
\be\label{dimnn}
N_{\mu_3\mu_4}^{\lambda_1}N_{\lambda_1\mu_1}^{\lambda_2}N_{\lambda_2\mu_2}^{\lambda_3}N_{\lambda_3\mu_5}^{\mu_6}\, .
\ee
 Recall that here $(\mu_3,\mu_4)$  are  labels of the Cardy state on the first  end of the strip, $(\mu_5,\mu_6)$  are  labels of the Cardy state on the second end of the strip,
and $(\mu_1,\mu_2)$ are the labels of the defect.

To interpret  this result let us remind some general facts on a semisimple 2D topological theory on a world-sheet with boundary \cite{Moore:2006dw}.
First of all let us recall that the whole content of the 2D topological field theory is encoded in a finite-dimensional commutative
Frobenius algebra ${\cal C}$. In the case when  ${\cal C}$ is semisimple it can be realized as the algebra of complex-valued functions
on a finite set $X={\rm Spec}\,{\cal C}$, which can be considered as a toy "space-time``.
Using sewing constraints of open topological theory  it was proved in   \cite{Moore:2006dw} that every boundary condition $a$ is
realized by a collections of vector spaces corresponding to each point of $X$: $x\rightarrow V_{x,a}$.
This can be considered as a vector bundle over finite space-time, in  agreement with the K-theory interpretation of boundary conditions.
The Hilbert space of open string with  boundary conditions specified by $a$ and $b$ is given by the bundle morphism:
\be
H_{ab}=\oplus_x{\rm Hom}(V_{x,a}; V_{x,b})\, .
\ee

Consider now an open topological $G/G$ coset. Note that in this case the points of $X$ are labelled by integrable primaries.
Let us remind first the situation without defect considered in \cite{Gawedzki:2001ye}.
The dimension of the Hilbert space for this case can be derived from (\ref{dimnn}) putting there $\mu_1$ and $\mu_2$ equal to vacuum state:
\be\label{dimnnn}
N_{\mu_3\mu_4}^{\lambda}N_{\lambda\mu_5}^{\mu_6}\, .
\ee
This can be interpreted  saying that the Hilbert space of the open string with the Cardy boundary conditions $(\mu_3,\mu_4)$  and $(\mu_5,\mu_6)$ at the ends
is
\be
H_{\mu_3,\mu_4;\mu_5,\mu_6}=\oplus_{\lambda}{\rm Hom}(W_{\mu_3\mu_4\lambda};W_{\mu_5\mu_6\lambda})\, ,
\ee
where $W_{\mu\nu\lambda}$ are spaces of three points conformal blocks.
This implies that the Cardy state $(\mu.\nu)$ is given by the vector bundle
\be
\lambda\rightarrow W_{\mu\nu\lambda}\, .
\ee

Now consider the case with a defect $(\mu_1,\mu_2)$.

It is well known  (see e.g.\cite{Petkova:2001zn,Graham:2003nc,Frohlich:2006ch,Kapustin:2006pk,Schweigert:2006af}), that open string propagating  with boundary conditions $a$ and $b$ with inserted defect $d$
can be considered, as propagating between one of the original boundary conditions, say $a$, and the second transformed by defect: $d\ast b$.
According to formula (\ref{dimnn}) the transformed state corresponds to the spaces $V_{\lambda,\mu_1,\mu_2,\mu_5,\mu_6}$ with the dimensions 
\be
N_{\lambda_1\mu_1}^{\lambda_2}N_{\lambda_2\mu_2}^{\lambda_3}N_{\lambda_3\mu_5}^{\mu_6}\, ,
\ee
and therefore can be considered as transformed by tensoring and summing with the space of 4-point conformal blocks  $W_{\mu_1\mu_2\lambda_1\lambda_3}$:
\be
V_{\lambda_1,\mu_1,\mu_2,\mu_5,\mu_6}=\oplus_{\lambda_3}W_{\mu_1\mu_2\lambda_1\lambda_3}\otimes W_{\mu_5\mu_6\lambda_3}\, .
\ee
 
This suggests  the following general description of defects in semisimple 2D TFT's. It seems that to every defect separating 2D TFT's with  "space-time``'s $X$
and $Y$
 corresponds a collection of spaces $V^D_{x,y}$ where $x\in X$ and $y\in Y$. This can be considered as a fibre bundle over $X\times Y$. 
Then the boundary condition given by the fibre bundle $V_{y}$ over $Y$   is transformed to the boundary condition corresponding to the following bundle over $X$:
\be\label{fourtr}
x\rightarrow \oplus_y V^D_{x,y}\otimes V_{y}\, .
\ee
It is interesting to note that the transformation (\ref{fourtr})  can be viewed as a discrete Fourier-Mukai transform 
in agreement with the general 
interpretation  of the defect worldvolume or bi-brane as kernel of the Fourier-Mukai transform suggested in \cite{Brunner:2008fa,Sarkissian:2008dq,Kapustin:2009av}.

Let us elaborate now on  fusion of defects.  For this purpose consider an open string with insertion of two defects.
The Hilbert space in this case is given by the space of 8-point conformal blocks. Along the same lines we conclude 
that the fusion of two defects $(\mu_1,\mu_2)$ and $(\nu_1,\nu_2)$ is given by the space of 6-point conformal blocks:
$W_{\mu_1,\mu_2,\nu_1,\nu_2,\lambda_1,\lambda_2}$. According  to the factorization properties of the space of conformal blocks
this space can be expressed through the space of 4-point conformal blocks:
\be
W_{\mu_1,\mu_2,\nu_1,\nu_2,\lambda_1,\lambda_2}=\oplus_{\gamma}W_{\mu_1,\mu_2,\lambda_1,\gamma}\otimes
W_{\nu_1,\nu_2,\lambda_2,\gamma}\, .
\ee

This suggests that in general the fusion of two defects given by  the bundles $V^{D_1}_{x,y}$ and $V^{D_2}_{y,z}$ over the spaces $X\times Y$ and $Y\times Z$
is given by the equation:
\be\label{deffus}
 V^{D_1\ast D_2}_{x,z}=\oplus_y V^{D_1}_{x,y}\otimes V^{D_2}_{y,z}\, .
\ee

It is interesting to note that equation (\ref{deffus})  appeared as a composition rule in the 2-category of matrices of vector spaces (see for example \cite{gk}).
The relation with 2-categories actually can be traced further.

Note that equation (\ref{dimdefcyl}) for the dimension of the $G/G$ theory on a cylinder with a defect can be written as the dimension
of the space $\sum_{\lambda} W_{\mu,\nu,\lambda,\lambda}$:
\be
{\rm dim}\, H_{d_{\mu,\nu}}={\rm dim} \sum_{\lambda} W_{\mu,\nu,\lambda,\lambda}\, .
\ee

We can conclude that probably in the general  case the dimension of the bulk theory with defect 
given by the collection of the spaces  $\{V_{x_1,x_2}, \;\; x_1,x_2\in X\}$,
is given by the dimension of the space $\oplus_{x}V^{D}_{x,x}$:
\be\label{cattrace}
{\rm dim}\, H_{d}={\rm dim}\, \oplus_{x}V^{D}_{x,x}\, .
\ee

The space  $\oplus_{x}V^{D}_{x,x}$ appears in \cite{gk} as categorical trace.

\vspace{0.3cm}

The study  of the conjectures (\ref{fourtr}), (\ref{deffus}) and (\ref{cattrace}) for general 2D semisimple TFT is left for future work.

\vspace{2cm}

\noindent {\bf Acknowledgements} \\[1pt]
I am grateful to Shmuel Elitzur, Christoph Schweigert,  Ingo Runkel, J\"urgen Fuchs and  Jens Fjelstad for useful discussions.

\newpage

\appendix

\section{Useful formulae}
In this appendix we collect some useful properties of the two-form $\Psi(h,g)$ defined by formula (\ref{formim}).
%\be
%\Psi(h,g)={\rm tr}\; h^{-1}d h(g^{-1}d g+d g g^{-1}+g^{-1}h^{-1}d h g)
%\ee

\be\label{om1}
\Psi(hL,p)=\Psi(L,p)+\Psi(h,LpL^{-1})\, .
\ee
\be\label{om2}
\Psi(Lh^{-1}, hph^{-1})=\Psi(L,p)-\Psi(h,p)\, .
\ee
\be\label{om}
\omega_{\mu}(hCh^{-1})-\omega_{\mu}(C)=-\Psi(h,C)\, .
\ee
%where $C=kfk^{-1}$.

%where $C_1=k_1f_1k_1^{-1}$,  $C_2=k_2f_2k_2^{-1}$
\be\label{om3}
\Omega^{(2)}(hC_1h^{-1},hC_2h^{-1})-\Omega^{(2)}(C_1,C_2)=-\Psi(h,C_1C_2)\, ,
\ee

\be\label{om5}
\Psi(h,C_1C_2)=\Psi(h,C_1)+\Psi(h,C_2)+(\tilde{C}_1^{-1}d\tilde{C}_1d\tilde{C}_2\tilde{C}_2^{-1}-C_1^{-1}dC_1dC_2C_2^{-1})\, ,
\ee
where $\tilde{C}_1=hC_1h^{-1}$  and $\tilde{C}_2=hC_2h^{-1}$\, .

\be\label{om4}
\omega^{WZW}(hgh^{-1})-\omega^{WZW}(g)=-d\Psi(h,g)\, .
\ee

\section{Defect Equations of motion}
Computing variation of the  action (\ref{actdefg})  one obtains for the defect part:
\bea\label{dffff}
{\rm tr}[g_1^{-1}\delta g_1(g_1^{-1}\partial_{+}g_1+g_1^{-1}\partial_{-}g_1)]dt
-{\rm tr}[g_2^{-1}\delta g_2(g^{-1}\partial_{+}g_2+g_2^{-1}\partial_{-}g)]dt\\ \nonumber
+2{\rm tr}[\delta g_1g_1^{-1}A_{1-}-A_{1+}g_1^{-1}\delta g_1-(\delta g_2g_2^{-1}A_{2-}-A_{2+}g_2^{-1}\delta g_2)]dt+B=0\, .
\eea
The last term $B$ is a one-form satisfying the relation:
\be\label{formb}
{\rm tr}(g_1^{-1}\delta g_1(g_1^{-1}d g_1)^2)-{\rm tr}(g_2^{-1}\delta g_2(g_2^{-1}d g_2)^2)-\delta\varpi=dB\, .
\ee
Recalling that the first two terms come from the equation
\be
\delta\omega^{WZ}=d[{\rm tr}(g^{-1}\delta g(g^{-1}d g)^2)]\, ,
\ee
we see that the existence of the one-form $B$ satisfying  (\ref{formb}) is a consequence of the equation  (\ref{extdef}).
Using (\ref{bibrane}) one can compute $B$ explicitly:
\bea\label{bexpo}
&&B=A_{\mu_1}(C_1)+A_{\mu_2}(C_2)+{\rm tr}[C_2^{-1}\delta C_2dC_1C_1^{-1}-\delta C_1C_1^{-1}C_2^{-1}dC_2-\\ \nonumber
&&\delta pp^{-1}(C_2C_1)^{-1}d(C_2C_1)
+(C_2C_1)^{-1}\delta(C_2C_1)dpp^{-1}-L^{-1}\delta Lp^{-1}dp+\\ \nonumber
&&p^{-1}\delta pL^{-1}dL-
L^{-1}\delta Ldpp^{-1}+\delta pp^{-1}L^{-1}dL
-L^{-1}\delta Lp^{-1}L^{-1}dLp+\\ \nonumber
&&L^{-1}\delta LpL^{-1}dLp^{-1}]\, .
\eea

The one-form $A_{\mu}(C)$ was defined in \cite{Elitzur:2000pq}:
\be
A_{\mu}(C)={\rm tr}[h^{-1}\delta h(f_{\mu}^{-1}h^{-1}dhf_{\mu}-f_{\mu}h^{-1}dhf_{\mu}^{-1})]\, ,
\ee

where $C=hf_{\mu}h^{-1}$, $f_{\mu}=e^{2\pi i\mu/k}$,

and satisfies:
\be
{\rm tr}(g^{-1}\delta g(g^{-1}d g)^2)|_{g=C}-\delta\omega_{\mu}(C)=dA_{\mu}(C)\, .
\ee

$A_{\mu}(C)$ satisfies also another important relation on the time-line:
\be\label{bndrmndr}
{\rm tr}[g^{-1}\delta g(g^{-1}\partial_{+}g+g^{-1}\partial_{-}g)]dt+A_{\mu}(C)={\rm tr}[2\delta hh^{-1}(\partial_{+}gg^{-1}-g^{-1}\partial_{-}g)]dt\, ,
\ee
where $g=C$.  Let us explain the meaning of this equation.

 The left hand side of the (\ref{bndrmndr}) is a particular case of  (\ref{dffff})
and describes boundary equation of motion of the WZW model with the boundary condition
specified by the conjugacy class $C$, while the right hand side proportional to $J_L+J_R$,
what is the condition for the diagonal chiral algebra preservation.

 Now, using  (\ref{bibrane}) and (\ref{bexpo}), one can show by a straightforwatd calculation, that (\ref{dffff})
implies  the equations (\ref{dee1}), (\ref{dee2}), (\ref{dee3}) in section 4.

\section{Symplectic forms of the moduli space of flat connections on a Riemann surface}

In this appendix we briefly review the symplectic phase space of the Chern-Simons theory on the three-dimensional manifold of
the form $M\times R$, where $M$ is two-dimensional Riemann surface, $R$ is time direction, with $n$ time-like
Wilson lines assigned with representations $\lambda_i$.
It was shown in \cite{Elitzur:1989nr,Witten:1988hf}  that the phase space of the Chern-Simons theory in such a situation 
is given by the moduli space of flat connections on the Riemann surface $M$  punctured at the points
$z_i$ where Wilson lines hit $M$,  with the holonomies around punctures belonging to the conjugacy classes
 $C_G^{\lambda_i}=\eta e^{2\pi i\lambda_i/k}\eta^{-1}$. Therefore denoting holonomies
around handles $a_j$ and $b_j$ by $A_j$ and $B_j$, and around punctures by $M_i\in  C_G^{\lambda_i}$ we arrive at the conclusion that the phase
space of the Chern-Simons theory is
\be
{\cal F}_{g,n}=G^{2g}\times \prod_{i=1}^n C_G^{\lambda_i}
\ee
subject to the relation
\be\label{relmod}
[B_{g}, A_{g}^{-1}]\cdots[B_{1}, A_{1}^{-1}]M_n\cdots M_1=I\, ,
\ee
where
\be
[B_j, A_j]=B_jA_jB_j^{-1}A_j^{-1}\, ,
\ee
and to the adjoint group action.

The symplectic form on ${\cal F}_{g,n}$ was derived in \cite{Alekseev:1993rj} and has the form:
\be\label{omms2}
\Omega_{{\cal M}_{g,n}}=\sum_{i=1}^n\Omega_{M_i} +\sum_{j=1}^{g}\Omega_{H_j}\, ,
\ee
where
\be
\Omega_{M_i}={k\over 4\pi} \omega_{\lambda_i}(M_i)
+{k\over 4\pi} {\rm tr}(K_{i-1}^{-1}\delta K_{i-1}K_{i}^{-1}\delta K_{i})\, ,
\ee
\bea
\Omega_{H_j}&=&{k\over 4\pi}\Psi(B_j,A_j)+{k\over 4\pi}({\rm tr}(K_{n+2j-2}^{-1}\delta K_{n+2j-2}K_{n+2j-1}^{-1}\delta K_{n+2j-1})\\ \nonumber
&+&
{\rm tr}(K_{n+2j-1}^{-1}\delta K_{n+2j-1}K_{n+2j}^{-1}\delta K_{n+2j}))\, ,
\eea
and where
\be
K_i=M_i\cdots M_1\hspace{1cm} i\leq n\, ,
\ee
\bea
K_{n+2j-1}=A_j[B_{j-1}, A_{j-1}^{-1}]\cdots[B_{1}, A_{1}^{-1}]K_n\, ,\\ \nonumber
K_{n+2j}=[B_{j}, A_{j}^{-1}]\cdots[B_{1}, A_{1}^{-1}]K_n\hspace{1cm} 1\leq j\leq g\, .
\eea
$K_0$ can be chosen to be equal to the unity element. According to (\ref{relmod})  also
\be
K_{n+2g}=I\, .
\ee
$\omega_{\lambda}(M)$ and $\Psi(B,A)$ are defined in equations (\ref{omfik}) and (\ref{formim}) correspondingly.

It was also proved in \cite{Alekseev:1993rj} that quantization of the moduli space ${\cal F}_{g,n}$
with the symplectic form (\ref{omms2}) leads to the space of $n$-point conformal blocks on a Riemann 
surface of the genus $g$.

The last piece of information which we need is the symplectic form on the moduli space of flat connections 
on the punctured sphere with holes $S^2_{n.m}$, where $n$ as before is number of punctures and $m$
is number of holes. It was argued in \cite{Elitzur:1989nr,Gawedzki:2001rm} that the corresponding  symplectic
form $\Omega_{S^2_{n,m}}$ is given by:
\be\label{decsn}
\Omega_{S^2_{n,m}}=\Omega_{S^2_{n+m,0}}+\sum_{i=1}^m\Omega_i^{\rm LG}\, ,
\ee
where $\Omega^{\rm LG}$ is defined in (\ref{omlg}) and its geometrical quantization 
leads to the integrable representation of the affine algebra $\hat{\rm g}$ at level $k$.

\end{document}